\documentclass[12pt,a4paper,notitlepage]{article}
\usepackage{geometry}
\usepackage{gfsbodoni}
\usepackage{xcolor} 
\usepackage{hyperref}
\usepackage[utf8]{inputenc}
\usepackage{amsmath}
\usepackage{mathtools,amssymb}
\usepackage{amsfonts}
\usepackage{amssymb}
\usepackage{relsize}
\usepackage{physics}
\usepackage{graphicx}
\graphicspath{{./figures/}}
\usepackage{fancyhdr}
\usepackage{cite}
\usepackage[noblocks]{authblk}
\usepackage[switch, modulo]{lineno}
\usepackage[margin=1cm]{caption}
\newcommand{\comment}[1]{}
\hypersetup{
colorlinks=true,
linkcolor=blue,
citecolor=magenta,
filecolor=magenta,
urlcolor=cyan
}
\title{\vspace{-1.8cm} \Large Spectral broadening from turbulence in multiscale lower hybrid current drive simulations}
\author[1]{\large Bodhi Biswas,\thanks{Present affiliation: University of York, UK. E-mail: \texttt{bodhi.biswas@york.ac.uk}}}
\author[1]{\large Paul Bonoli}
\author[1]{\large Abhay Ram}
\author[1]{\large Anne White\vspace{-1ex}}
\affil[1]{\normalsize Plasma Science and Fusion Center, Massachusetts Institute of Technology\vspace{-3ex}}
\date{\normalsize \today \vspace{-4ex}}

\begin{document}
\maketitle
\begin{abstract}
\emph{The scattering of lower hybrid (LH) waves due to scrape-off layer (SOL) filaments is investigated. It is revealed that scattering can account for the LH spectral gap without any ad hoc modification to the wave-spectrum. This is shown using a multiscale simulation approach which allows, for the first time, the inclusion of full-wave scattering physics in ray-tracing/Fokker-Planck calculations. In this approach, full-wave scattering probabilities are calculated for a wave interacting with a statistical ensemble of filaments. These probabilities are coupled to ray-tracing equations using radiative transfer (RT) theory. This allows the modeling of scattering along the entire ray-trajectory, which can be important in the multi-pass regime. Simulations are conducted for lower hybrid current drive (LHCD) in Alcator C-Mod, resulting in excellent agreement with experimental current and hard X-ray (HXR) profiles. A region in filament parameter space is identified in which the impact of scattering on LHCD is saturated. Such a state coincides with experimental LHCD measurements, suggesting saturation indeed occurs in C-Mod, and therefore the exact statistical properties of the filaments are not important.}
\end{abstract}

\section{Introduction}
Lower hybrid waves are an efficient means to non-inductively drive current in a tokamak via electron Landau damping (ELD) \cite{fischTheoryCurrentDrive1987}. It is an attractive actuator for current profile shaping, which has been successfully demonstrated in several tokamaks\cite{ideLHCDCurrentProfile2000, peyssonHighPowerLower2001,litaudonFullyNoninductiveCurrent2002,pericoli-ridolfiniHighPlasmaDensity1999,shiraiwaDesignInitialExperiment2011}. There is also interest in targeted LHCD for neoclassical tearing mode suppression \cite{lahayeNeoclassicalTearingModes2006,reimanSuppressionTearingModes2018a}.

The condition for strong linear ELD is \cite{bonoliLinearTheoryLower1984}
\begin{equation}
\abs{N_{||}} \geq N_{||,\text{ELD}} \equiv \frac{c}{3 v_{te}} 
\end{equation}
where $N_{||} \equiv \bold{N} \cdot \bold{B}/B$ is the parallel refractive index (with respect to the background magnetic field $\bold{B}$), $c$ is the speed of light, and $v_{te}=\sqrt{2 T_{e}/m_e}$ is electron thermal velocity (with $T_{e}$ and $m_{e}$ being the electron temperature and mass, respectively). If the initial parallel refractive index $\abs{N_{||0}} <  N_{\text{ELD}}$, the wave is expected to first undergo $|N_{||}|$-upshift due to toroidicity until the condition in Eq. (1) is met \cite{bonoliToroidalScatteringEffects1982}. For a sufficiently large \emph{spectral-gap} between $\abs{N_{||0}}$ and $ N_{\text{ELD}}$, the wave may require multiple passes through the plasma before this gap is bridged. This scenario is typical of present-day tokamaks due to low core $T_{e}$. 

\sloppy{A large spectral-gap poses difficulties for theory and modeling of LHCD. Ray-tracing/Fokker-Planck modeling of LHCD reveal large discrepancies between simulations and experiments in several tokamaks \cite{mumgaardLowerHybridCurrent2015,peyssonProgressHighpowerLower2000,garofaloDevelopmentHighPoloidal2017, peyssonLowerHybridCurrent2020a}. In high aspect ratio devices ($\mathcal{A} \equiv R_0/a \gtrsim 5$, where $R_0$ is the major radius, and $a$ is the minor radius), like WEST and TRIAM-1, toroidicity is insufficient to bridge the spectral-gap via $\abs{N_{||}}$-upshift. In moderate aspect ratio devices ($\mathcal{A} \approx 3$), like Alcator C-Mod and EAST, toroidicity can accurately model the \emph{total} current drive efficiency, but not the radial current or HXR profiles \cite{mumgaardLowerHybridCurrent2015,garofaloDevelopmentHighPoloidal2017}. For example, in C-Mod, the simulated LH current profiles are not smooth and are peaked far off-axis, while experimental motional Stark effect (MSE) and HXR measurements suggest smooth profiles robustly peaked on-axis \cite{mumgaardLowerHybridCurrent2015}. Consequently, $N_{||}$-upshift due to toroidicity is not adequate for understanding experimental observations. There exist other contributing factors that must be taken into account in theoretical models.}

Several mechanisms have been proposed as possible explanations for the spectral gap. Ad hoc modifications to the launched LH wave-spectrum reveal that either $N_{||}$ broadening \cite{peyssonAdvancesModelingLower2016} or angle-broadening of the perpendicular refractive index ($\bold{N}_{\perp}$) \cite{baekModelInvestigationImpact2021} can explain experimental current drive results in C-Mod. 

The two theorized causes of $N_{||}$ broadening are parametric decay instabilities (PDI) \cite{porkolabTheoryParametricInstability1974a,cesarioModelingLowerHybridCurrent2004} and scattering from parallel density fluctuations in front of the LH antenna \cite{madiPropagationLowerHybrid2015}. In FTU, a combination of modeling and experiment shows PDI is likely responsible for low LHCD efficiency at high densities \cite{cesarioSpectralBroadeningParametric2014}. However, there is little theoretical support for - or experimental evidence of - strong PDI in low and moderate density C-Mod discharges ($\bar{n}_{e} < 10^{20}\, \text{m}^{-3}$) where the spectral gap persists \cite{baekHighDensityLHRF2015}. Likewise, there is little evidence of density fluctuations with the large parallel gradients required to induce a significant $N_{||}$-broadening. 

The most likely mechanism for $\bold{N}_{\perp}$ angle-broadening is scattering from turbulent scrape-off layer (SOL) fluctuations. Prior models have employed either ray-tracing \cite{bellanEffectDensityFluctuations1978,peyssonRFCurrentDrive2011} or wave-kinetic treatments \cite{bonoliToroidalScatteringEffects1982,andrewsScatteringLowerhybridWaves1983,bertelliEffectsScatteringEdge2013} for scattering in drift-wave turbulence. These models have demonstrated modified current profiles, but cannot match experimental observations. Gas-puff imaging (GPI) \cite{zwebenEdgeTurbulenceImaging2002} and statistical analysis of Langmuir probe measurements \cite{gravesSelfsimilarDensityTurbulence2005} in the SOL have motivated the modeling of LH scattering from intermittent, field-aligned filaments. The extent of $\bold{N}_{\perp}$ angle-broadening from filaments is greater than in ``equivalent'' drift-wave turbulence \cite{biswasStudyTurbulenceinducedRefraction2020}. A recent hybrid full-wave/statistical model for wave-filament interactions was developed to model the modification to the LH wave-spectrum in front of the antenna \cite{biswasHybridFullwaveMarkov2021}. Multiple wave-filament interactions are accounted for using the radiative transfer (RT) equation. This treatment allows the modeling of realistic turbulence parameters without being restricted to the validity constraints of the ray-tracing or random phase approximation. The study finds a large angle-broadening of the incident wave-spectrum, enough to robustly direct a fraction of LH power to damp on-axis on first pass through the plasma. In turn, the LH current profile is monotonic and peaked on-axis, in much closer agreement with experiment. In addition, asymmetric scattering of the LH wave is observed in angle-space. This is a full-wave effect only possible with intermittent density fluctuations, and therefore is not accounted for in prior ray-tracing or wave-kinetic treatments.

The hybrid full-wave/statistical scattering model discussed above is limited to a slab geometry with homogeneous background and turbulence parameters. It also only treats slow wave to slow wave ($S \rightarrow S$) scattering and ignores the fast ($F$) wave. Therefore this model can only approximately treat scattering directly in front of the LH antenna. In this paper, the more general RT equation is solved using a multiscale full-wave/ray-tracing solver. It allows the modeling of arbitrary geometry and both like-mode ($S\rightarrow S$, $F \rightarrow F$) and unlike-mode ($S\rightarrow F$, $F \rightarrow S$) scatter. An arbitrary geometry allows accounting for realistically tapered SOL turbulence profiles in a tokamak, and models scattering along the entire ray-trajectory (important in the multi-pass regime). The inclusion of all scattering modes is especially important near the mode-conversion density where like- and unlike-mode scattering probabilities are comparable. 

This multiscale model is applied to Alcator C-Mod, allowing, for the first time, the inclusion of self-consistent full-wave scattering physics in LHCD simulations. The resulting current and HXR profiles provide excellent matches to experimental measurements. Thus, the spectral-gap in C-Mod is resolved via turbulent scattering, and without any ad hoc modification to the wave-spectrum.

This paper is organized as follows. Section 2 will describe the multiscale scattering model. Section 3 discusses its application to LHCD in Alcator C-Mod discharges. Section 4 provides a discussion and summary of the findings.

\section{Multiscale scattering model}
From a modeling perspective, the LH wave is in an interesting and challenging range of wavelengths. The wavelength is small enough to employ ray-tracing in the quiescent core \cite{frankVerifyingRaytracingFokkerPlanck2022}, but large enough that full-wave modeling is required for common SOL turbulence parameters. On the other hand, the LH wavelength is small enough such that whole-device full-wave modeling in a turbulent plasma is prohibitively expensive. Therefore, a multiscale model that employs ray-tracing in the core and a full-wave solver in the SOL is a promising concept. The implementation is as follows. A Mie-scattering approach efficiently calculates the scattering probabilities for an incident plane wave interacting with a filament. The probabilities are averaged over a statistical ensemble of filament parameters. Next, the scattering of RF power in phase-space is modeled using the RT approximation. The RT equation is solved in a ray-tracing solver, where the scattering terms act as stochastic kicks to the ray-trajectory. Finally, the rays are used to calculate current drive using the quasi-linear Fokker-Planck equation.

\subsection{Review of single wave-filament interaction}
The Mie-scattering model for a single wave-filament interaction is briefly reviewed.  An incident LH plane wave, either the slow ($S$) or fast ($F$) mode, travels through a cold magnetized plasma with a homogeneous background density $n_0$ and magnetic field $\bold{B}$. The magnetic field is aligned along $\bold{\hat{e}}_{z}$, such that the parallel wave-number $k_{||}=k_{z}$. The wave trajectory is aligned such that $\bold{v}_{gr\perp} = v_{gr\perp} \bold{\hat{e}}_{x}$, where $\bold{v}_{gr\perp}$ is the perpendicular group velocity and $v_{gr\perp}>0$. An infinitely long, cylindrical, field-aligned filament passes through the origin with density $n_b$. Given this is a poloidally symmetric system (where $\theta = \text{tan}^{-1}(y/x)$), the Jacobi-Anger expansion is used to write the electric field in cylindrical coordinates ($\rho,\theta,z$). Consequently, the field everywhere is a series solution in poloidal mode-numbers ($m \equiv k_{\theta} \rho$),
\begin{subequations}
\begin{align}
E_{j \beta} & = e^{i(k_{||}z-\omega t)}\sum_{m=-\infty}^{+\infty} E_{jm} W_{j \beta m} e^{i m \theta}; \quad \quad\beta = \rho,\theta,z \\
W_{j\rho m} & = \xi_{jx} J_{m}^{'}(k_{j \perp}\rho) - i\xi_{jy} \frac{m}{k_{j \perp} \rho} J_{m}(k_{j \perp}\rho) \\
W_{j\theta m} & = i\xi_{jx} \frac{m}{k_{j \perp} \rho} J_{m}(k_{j \perp}\rho) + \xi_{jy} J_{m}^{'}(k_{j \perp}\rho) \\
W_{jzm} & = i \xi_{jz}J_{m}(k_{j \perp}\rho)
\end{align}
\end{subequations}
where $j=0,...,4$ is the wave index, indicating (0) the incident wave, (1,2) the slow/fast mode inside the filament, and (3,4) the slow/fast mode outside the filament. $\bar{\xi}_j = \{\xi_{jx},\xi_{jy},\xi_{jz}\}$ is the plane-wave polarization of wave $j$. $J_{m}$ is the Bessel function of the first kind and order $m$. $J_{m}^{'}$ is the first derivative of $J_{m}$ with respect to its argument. The coefficients $E_{jm}$ for the incident plane wave ($j=0$) are known through the Jacobi-Anger expansion. The coefficients for the remaining waves are found by imposing Maxwell's boundary conditions at the filament perimeter. This is detailed in Ram \emph{et al.}, (2016) \cite{ramScatteringRadioFrequency2016}.

This technique can be extended to a radially varying filament density profile, as detailed in Biswas \emph{et al.}, (2021) \cite{biswasHybridFullwaveMarkov2021}. A similar method was used to model ion-cyclotron wave scattering \cite{zhangInteractionFilamentsICRF2021}. For the purposes of this work, it is assumed the filament has a Gaussian density profile such that
\begin{equation}
n(\rho) -n_0= n_0 \left( \frac{n_b}{n_0} -1 \right) e^{-\left(\frac{2\sqrt{\text{ln}(2)}\rho}{a_b}\right)^{2}}
\end{equation}
where $n$ is the density, $n_b/n_0$ is the relative density of the filament, and $a_b$ is the full-width half max of the filament radial profile.

Following \cite{myraScatteringRadioFrequency2010,biswasHybridFullwaveMarkov2021}, the differential scattering-width is calculated by evaluating the radially scattered Poynting flux in the far-field limit
\begin{equation}
\sigma_{j \rightarrow j'}(\theta) = \frac{ \mp 2}{\pi} \frac{|\xi_{j'y}|^2 + |\xi_{j'z}|^2 + \frac{k_z}{k_{j'\perp}} \Re{\xi_{j'x} \xi_{j'z}^{*}} }{k_{jx} (|\xi_{jy}|^2 + |\xi_{jz}|^2) -k_{z} \Re{\xi_{jx} \xi_{jz}^{*}}} \abs{\sum_{m=-\infty}^{+\infty}i^{\pm m} E_{j'm} e^{i m \theta}}^{2}
\end{equation}
where now $j=S,F$ is the incident mode, $j'$ is the scattered mode, $\xi_{j\ell}$ is the $\ell$-component of the electric field polarization of mode $j$, and $k_{j\ell}$ is the $\ell$-component of the wave-number of mode $j$. The sign $\mp$ depends on the scattered mode. The differential scattering-width is analogous to a differential scattering cross-section. Likewise, $\sigma = \int \sigma (\theta) d\theta$ is the scattering-width, which is analogous to a scattering cross-section. Lastly, $\hat{\sigma} (\theta) \equiv \sigma (\theta)/\sigma$ is the normalized differential scattering-width and will be useful in writing the RT equation.

\subsection{Scattering through statistical ensemble of filaments}
To account for the statistical variation in filament relative density ($n_b/n_0$) and radial width ($a_b$), a joint probability distribution function $p(n_b/n_0, a_b)$ is defined for an ensemble of filaments. The quantities $\langle n_b/n_0 \rangle$ and $\langle a_b \rangle$ denote the average relative density and radial width of filaments. The average, or ``effective'', differential scattering-width is
\begin{equation}
\sigma_{\text{eff},j\rightarrow j'}(\theta) = \langle \sigma_{j\rightarrow j'}(\theta) \rangle = \int_{0}^{\infty} \text{d}a_{b} \int_{0}^{\infty}\text{d}\left(n_{b}/n_0\right) \, \sigma_{j \rightarrow j'}(\theta; n_b/n_0, a_b) p(n_b/n_0,a_b)
\end{equation}

A wave packet traveling in a straight line will encounter, on average, $\frac{f_p}{\pi \langle a_b \rangle^{2}}$ filaments per unit length in the perpendicular plane. Thus, the inverse mean-free-path to scatter is
\begin{equation}
\Sigma_{\text{eff},j\rightarrow j'} = \frac{f_p}{\pi \langle a_b \rangle^{2}}\sigma_{\text{eff},j\rightarrow j'}
\end{equation}
where $f_p$ is the packing-fraction, which is defined as the fractional area (in the perpendicular plane) inhabited by filaments. Note that $\Sigma_{\text{eff},j\rightarrow j'}$ and $\sigma_{\text{eff},j\rightarrow j'}(\theta)$ are dependent on the local plasma parameters, the incident wave's frequency and wavenumber, and the filament PDF. These functional dependencies are only stated explicitly when needed.

\subsection{Radiative transfer approximation}
Radiative transfer (RT) theory models the RF wave-spectrum as a wave-packet distribution function in phase-space, akin to the Fokker-Planck equation for a particle distribution function \cite{andrewsScatteringLowerhybridWaves1983}. We define $P_j(\bold{r},\bold{k})$ as the distribution corresponding to the LH mode $j$. The RT equation governing $P_j$ is
\begin{equation}
\left(\frac{dP_{j}}{dt}\right)_{\text{r}} + 2 \gamma(\bold{k}_{j},\bold{r})P_{j} = \left(\frac{dP_{j}}{dt}\right)_{\text{sct}}
\end{equation}
The $(...)_{\text{r}}$ term is the convective term accounting for the trajectory for a wave-packet/ray. The second term accounts for wave damping, where $\gamma$ is the damping rate. The left hand side (LHS) is routinely solved using ray-tracing/Fokker-Planck codes. The right hand term accounts for the added effect of scattering,
\begin{equation}
\begin{split}
\left(\frac{dP_{j}}{dt}\right)_{\text{sct}} & = \mathlarger{\sum}_{j'=S,F}  -\Sigma_{\text{eff},j \rightarrow j'}(k_{||},\bold{r}) |\bold{v}_{gr \perp}(\bold{k}_{j},\bold{r})| P_{j} \\
& + \mathlarger{\sum}_{j'=S,F} \Sigma_{\text{eff},j' \rightarrow j}(k'_{||},\bold{r}) |\bold{v}_{gr \perp}(\bold{k}_{j'},\bold{r})| \int_{-\pi}^{\pi}\hat{\sigma}_{\text{eff},j' \rightarrow j}(\chi-\chi';k_{||},\bold{r})P_{j'}(\chi',k_{||},\bold{r})d\chi'
\end{split}
\end{equation}
The transformation of $\theta \rightarrow \chi$ must first be explained. The variable $\theta$ describes the angle of scatter in the frame of a single filament-wave interaction, as described in Section 2.1. In contrast, $\chi$ is defined through $\bold{b} \cdot (\hat{\bold{e}}_{\psi} \times \bold{k}_{\perp}) = k_{\perp} \sin{\chi}$, where $\hat{\bold{e}}_{\psi}$ is the unit vector normal to the flux surface and directed outwards, and $\bold{b} \equiv \bold{B}/B$ is the unit vector along the magnetic field. Thus, by using $\chi$, the angular orientation of $\bold{k}_{\perp}$ is unambiguously defined in a tokamak geometry. The functional dependence of $P_j(\bold{r},\bold{k})$ can be mapped to $P_j(\chi,k_{||},\bold{r})$ and vice-versa using the dispersion relation for the appropriate mode, and the local magnetic geometry.

The two RHS terms in Eq. (8) account for angular ($\chi$) rotation of the perpendicular wave-vector $\bold{k}_{\perp}$ due to a wave-filament interaction.  The first term accounts for \emph{out-scatter} of $P_j$ from a phase-space volume element centered at $\chi$. The second term accounts for \emph{in-scatter} into this volume element from all other angles $\chi '$. Note that the scattering probabilities and background plasma parameters are allowed to vary in real-space. In addition, the summation over the slow and fast modes now ensure that both like-mode and unlike-mode scatter ($S \rightarrow F, F \rightarrow S$) are accounted for. In contrast, the system studied in \cite{biswasHybridFullwaveMarkov2021} was restricted to homogeneous scattering and background parameters, and neglected mode-conversion.

It should be clarified that, in generating scattering probabilities using a full-wave formalism, interference effects during a single wave-filament interaction are accounted for. However, in using the RT approximation, the interference effects of simultaneous \emph{multi}-filament scattering are ignored. This is a reasonable approximation as long as $k_{\perp}d \gg 1$, where $d$ is the average distance between filaments \cite{mishchenkoElectromagneticScatteringParticles2014}. Code comparison with a fully numeric full-wave solver finds that this multiscale method generally overestimates the effects of scattering for $f_p \gtrsim 0.15$ \cite{biswasHybridFullwaveMarkov2021}. Nevertheless, the multiscale model retains many important full-wave effects and is therefore a significant improvement over prior reduced RF-turbulence scattering models in the SOL.

\subsection{Coupling to ray-tracing code GENRAY}
The initial wave-spectrum, launched from the LH antenna, is discretized into rays. Each ray-trajectory is evolved using the ray-tracing equations in GENRAY \cite{smirnovGENRAYRayTracing2001}. The quasi-linear calculation of $\gamma$, accounting for Landau damping and collisions, is accomplished with the Fokker-Planck solver CQL3D \cite{harveyCQL3DFokkerPlanckCode2005}. In this way, the usual ray-tracing/Fokker-Planck calculation solves for the LHS of Eq. (7). 

The right hand side (RHS) of Eq. (7) introduces stochastic kicks to the ray-trajectories via a Monte Carlo process. Previous ray-tracing Monte Carlo scattering models exist for drift-wave turbulence \cite{bonoliToroidalScatteringEffects1982, bertelliEffectsScatteringEdge2013}. In GENRAY, after one time-step of $\Delta t$, the ray-trajectory is evolved from $\bold{r}$ to $\bold{r} + \Delta\bold{r}$, where $\bold{r}$ is a spatial coordinate and $\Delta \bold{r} = \bold{v}_{gr}\Delta t$. Given $\Sigma_{\text{eff},j\rightarrow j}$ is the inverse mean free path for like-mode scatter, the probability of this scattering event during this time-step is $p_{\text{sct}} = 1 - e^{-\Sigma_{\text{eff},j\rightarrow j} |\Delta \bold{r}_{\perp}|}\approx\Sigma_{\text{eff},j\rightarrow j} |\Delta \bold{r}_{\perp}|$, where $|\Delta \bold{r}_{\perp}|$ is the distance travelled perpendicular to $\bold{b}$. The approximation holds for $\Sigma_{\text{eff},j\rightarrow j} |\Delta \bold{r}_{\perp}| \ll 1$. The probability of unlike-mode scatter, and no scatter, are calculated in a similar manner. After each time-step, a uniformly distributed random variable $X_{1} \in [0,1]$ is sampled. Assuming the incident ray is a slow wave, there are three possibilities,
\begin{enumerate}
\item $X_{1} < \Sigma_{\text{eff},S \rightarrow S} |\Delta \bold{r}_{\perp}|$. Then a $S \rightarrow S$ scatter occurs.
\item $\Sigma_{\text{eff},S \rightarrow S} |\Delta \bold{r}_{\perp}| \leq X_{1} < (\Sigma_{\text{eff},S \rightarrow S}+\Sigma_{\text{eff},S \rightarrow F}) |\Delta \bold{r}_{\perp}|$. Then a $S \rightarrow F$ scatter occurs.
\item $X_{1} \geq (\Sigma_{\text{eff},S \rightarrow S}+\Sigma_{\text{eff},S \rightarrow F}) |\Delta \bold{r}_{\perp}|$. Then no scatter occurs.
\end{enumerate}
The above stochastic scattering procedure is only valid when the scattering probability between time-steps is small ($(\Sigma_{\text{eff},S \rightarrow S}+\Sigma_{\text{eff},S \rightarrow F}) |\Delta \bold{r}_{\perp}| \ll 1$). Once a scattering event is initiated, another uniformly distributed random variable $X_{2} \in [0,1]$ is sampled. A new variable is introduced:
\begin{equation}
\tilde{\sigma}_{\text{eff},j\rightarrow j'}(\chi) = \frac{1}{\sigma_{\text{eff},j\rightarrow j'}} \int_{-\pi}^{\chi} \sigma_{\text{eff},j \rightarrow j'}(\chi')d\chi'
\end{equation}
where $\tilde{\sigma}_{\text{eff},j\rightarrow j'}(\Delta \chi)$ is the cumulative distribution function (CDF) of rotating the incident ray $\bold{k}_{\perp}$ by angle $\Delta \chi$. Inversely,
\begin{equation}
\Delta \chi = \tilde{\sigma}_{\text{eff},j\rightarrow j'}^{-1}(X_{2})
\end{equation}
 Therefore, the incident ray is converted to a scattered ray, by first undergoing mode-conversion in the case of unlike-mode scatter, and then rotating $\bold{v}_{gr \perp}$ by $\Delta \chi$ with respect to $\bold{b}$. Note that, in the case of like-mode scatter, this is equivalent to rotating $\bold{k}_{\perp}$ by $\Delta \chi$. In the case of unlike-mode scatter, the direction of $\bold{k}_{\perp}$ is rotated by $\Delta \chi + \pi$ because the fast wave is forward-propagating ($\bold{k}_{\perp} \cdot \bold{v}_{gr\perp} = |k_{\perp} v_{gr\perp}|$) while the slow wave is backward-propagating ($\bold{k}_{\perp} \cdot \bold{v}_{gr\perp} = -|k_{\perp} v_{gr\perp}|$).
 
The values of $\Sigma_{\text{eff},j \rightarrow j'}$ and $\tilde{\sigma}_{\text{eff},j\rightarrow j'}(\chi)$ depend on $k_{||}$ and $\bold{r}$. They are implemented in GENRAY as lookup tables parameterized by $k_{||}$, $\psi$, and $\theta_{p}$ (where $\psi$ is the normalized radial coordinate and $\theta_{p}$ is the poloidal coordinate typically used in tokamaks). The filament statistics (and therefore the scattering probabilities) are made to vary such that relative density fluctuations increase with outward distance from the separatrix. Due to the ballooning nature of SOL turbulence, fluctuations are smaller on the high field side. This is accounted for via the $\theta_{p}$ dependence of the scattering probabilities. For simplicity, $f_p$ is assumed constant (though it can also be made to vary spatially). Details on the radial and poloidal tapering of scattering probabilities can be found in Appendix A.

\section{Application to Alactor C-Mod}
A well-studied \cite{mumgaardLowerHybridCurrent2015,biswasStudyTurbulenceinducedRefraction2020,baekModelInvestigationImpact2021,biswasHybridFullwaveMarkov2021} low-$\bar{n}_e$, fully non-inductive discharge (\#1101104011) is modeled in GENRAY/CQL3D. Both HXR and MSE measurements are available, allowing for direct comparison to the simulated current density HXR count profiles. In modelling a quasi-steady-state time-slice at which $V_{\text{loop}}\approx 0$, it can be safely assumed that Ohmic contribution to the current profile is negligible.

Smooth electron density and temperature profiles are fit to Thompson scattering measurements. A kinetic EFIT profile is used (constrained by MSE measurements and produced in \cite{mumgaardLowerHybridCurrent2015}). To model the SOL geometry, the two-point model is used to generate $n,T$ profiles outside the separatrix \cite{shiraiwaImpactSOLPlasma2015}. It is assumed that the SOL e-folding width $\lambda_{\text{SOL}} = 7\,$mm, and that $Z_{\text{eff}}=1.5$ everywhere.

\subsection{Scattering probabilities}
The filament joint-PDF $p(n_b/n_0,a_b)$ is assumed to be a bivariate normal distribution. Relevant parameters include the average ($\langle ... \rangle$), standard deviation ($s$), and skewness ($\Gamma$) for $n_b/n_0$ and $a_b$. (See Appendix A for a complete definition of these parameters, and how they vary radially and poloidally. Figures \ref{fig:app1} and \ref{fig:app2} summarize the spatial variations of the filament PDF). 

\begin{figure}[!h]
\centering
\includegraphics[width=17cm, height=10cm]{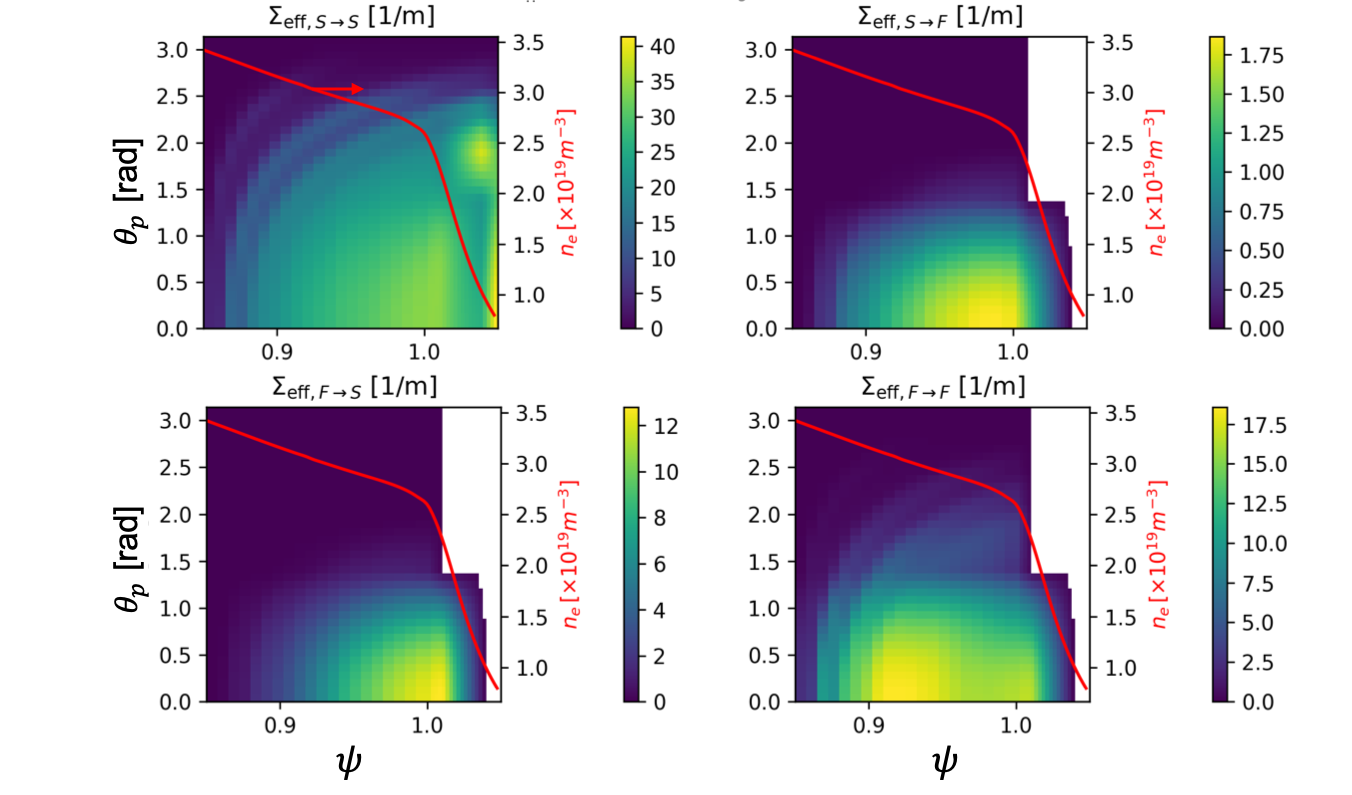}
\caption[font=5]{$\Sigma_{\text{eff},i\rightarrow j}(\psi,\theta_{p},N_{||}=1.6)$ plotted for $\langle n_{0}/n_{b}\rangle_{\text{grill}} = 3$, $\langle a_{b} \rangle=0.5\,$cm, $s_{a_b} = 0.2\,$cm, $\Gamma_{a_b} = 7$, and $f_{p}$=0.2. The white regions correspond to an evanescent FW. The red line indicates background density.}
\label{fig:5.9}
\end{figure}

For the purposes of illustrating how scattering probabilities vary in the SOL, it is assumed that $\langle n_{0}/n_{b}\rangle_{\text{grill}} = 3$, $\langle a_{b} \rangle=0.5\,$cm, $s_{a_b} = 0.2\,$cm, and $\Gamma_{a_b} = 7$. (The parameter $\langle n_{0}/n_{b}\rangle_{\text{grill}}$ sets  $\langle n_{0}/n_{b}\rangle$ at the LH antenna.) Figures \ref{fig:5.9} plot $\Sigma_{\text{eff},i\rightarrow j}(\psi,\theta_{p})$ for $N_{||}=1.6$ and $f_{p}=0.2$ (though one can argue that so high a $f_p$ is not strictly valid given the discussion in Section 2.3). As expected, scattering is strongest in the far-SOL at the outer mid-plane, where $\Sigma_{\text{eff},S\rightarrow F}\approx 40 \,\text{m}^{-1}$. Conversely, scattering is weakest further into the core and closer to the inner mid-plane. Striations and local maxima in these plots, most notably for $\Sigma_{\text{eff},S\rightarrow S}$ and $\Sigma_{\text{eff},F\rightarrow F}$, are evidence of resonances akin to those found in Mie-scattering.

It should be noted that, while $S\rightarrow F$ and $F\rightarrow S$ are ``opposite'' scattering events, their probabilities are not equal. Within the range of parameters tested, it is most common that $\Sigma_{\text{eff}, F \rightarrow S} > \Sigma_{\text{eff}, S \rightarrow F}$. In other words, there is preferential scattering from $S \rightarrow F$.

\begin{figure}[!h]
\centering
\includegraphics[width=17cm, height=10cm]{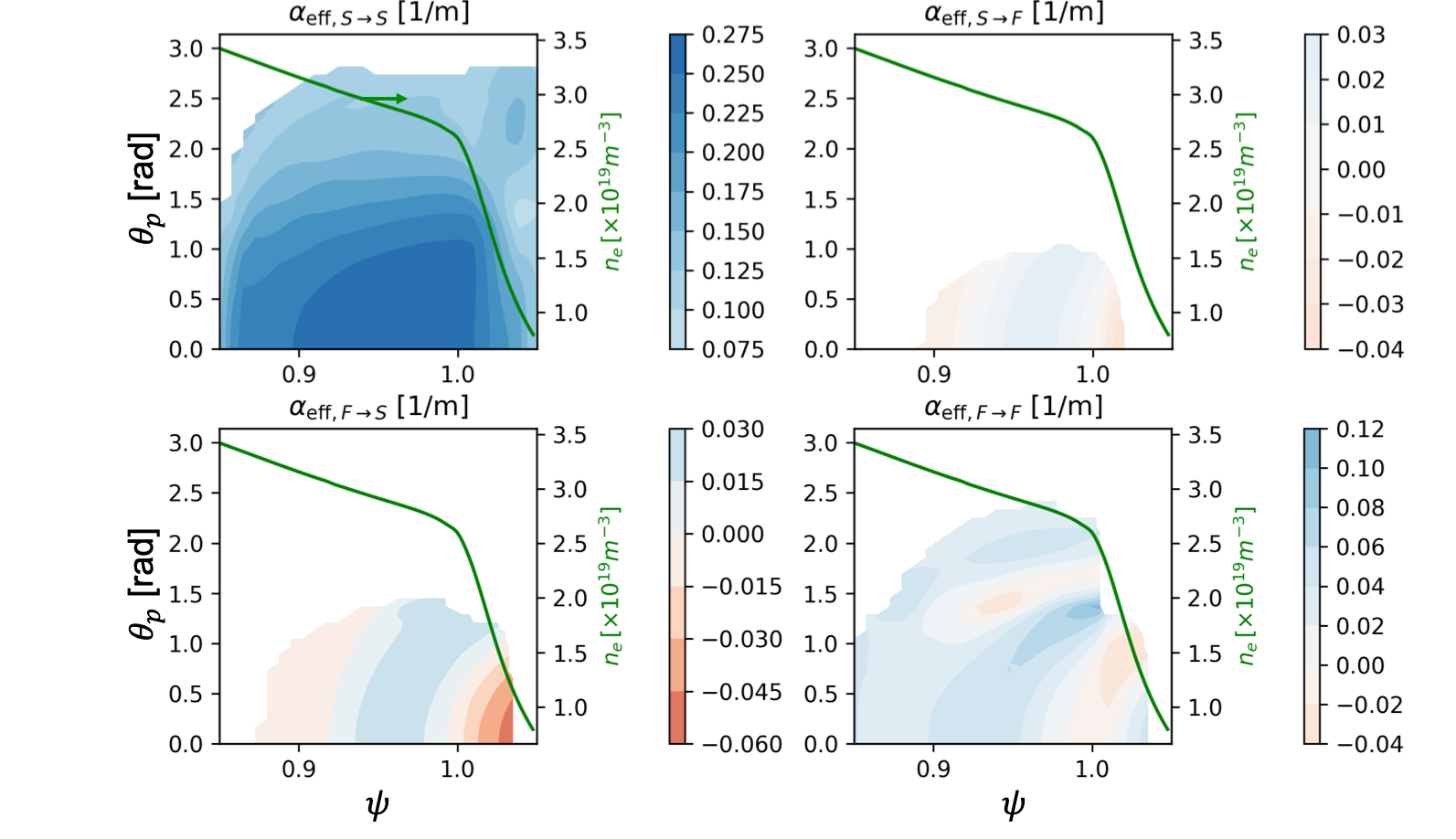}
\caption[font=5]{Asymmetric scattering metric $\alpha_{\text{eff},i\rightarrow j}(\psi,\theta_{p},N_{||}=1.6)$ plotted for same filament joint-PDF's as in Figure \ref{fig:5.9}. White regions correspond to $\Sigma_{\text{eff},i \rightarrow j} < 1 \,\text{m}^{-1}$, where scattering is negligible. The green line indicates background density. }
\label{fig:5.11}
\end{figure}

Asymmetric scatter (in angle space) is a full-wave effect, and is found to make a significant modification to the LH wave-spectrum. The physical explanation for asymmetric scatter is discussed in \cite{biswasHybridFullwaveMarkov2021}. Figures \ref{fig:5.11} plots the effective asymmetric scatter metric $\alpha_{\text{eff},i \rightarrow j}$, defined as:
\begin{equation}
\alpha_{\text{eff},j\rightarrow j'} \equiv \int_{0}^{\pi} \langle \, \hat{\sigma}_{j \rightarrow j'}(\chi) \rangle \text{d} \chi -\frac{1}{2}
\end{equation}
An $\alpha_{\text{eff},j\rightarrow j'} > 0$ ($<0$) corresponds to scatter from mode $j$ to $j'$ preferentially in the $+\chi$ ($-\chi$) direction. Since LH waves are generally launched with $N_{||}<0$ in order to drive co-current, it can be shown that preferential $+\chi$ rotation equates to preferential scatter \emph{away} from the plasma core. Conversely, $-\chi$ rotation leads to scattering \emph{towards} the core. Figures \ref{fig:5.9} and \ref{fig:5.11} show the strongest asymmetry is from $S \rightarrow S$ scatter (in the sense that $\Sigma_{\text{eff},S \rightarrow S}$ is rather large and $|\alpha_{\text{eff},S \rightarrow F}|$ is robustly positive). Thus, there is a net scatter of power away from the core everywhere in the SOL. Although not shown here, this trend is insensitive to $N_{||}$ in the ranges expected in C-Mod.

\subsection{Parametric scan of turbulence parameters}
As a metric for simulation accuracy compared to experiment, we define the \emph{normalized mean squared error} for an arbitrary simulated measurement $X$ as
\begin{equation}
\bar{S}^{2} = \frac{|| X_{\text{sim}} - X_{\text{exp}} ||}{|| X_{\text{simref}} - X_{\text{exp}}||}
\end{equation}
where ``sim'' denotes the simulated values, ``exp'' the experimental values, and ``simref'' the simulated values in the reference (no turbulence) case. Further, we define $\bar{S}^{2}_{J_{\phi}}$ as replacing the array $X$ in Eq. (12) with toroidal current density ($J_{\phi}$) evaluated at the radial grid points modeled in CQL3D.

\begin{figure}[!h]
\centering
\includegraphics[width=16cm, height=11cm]{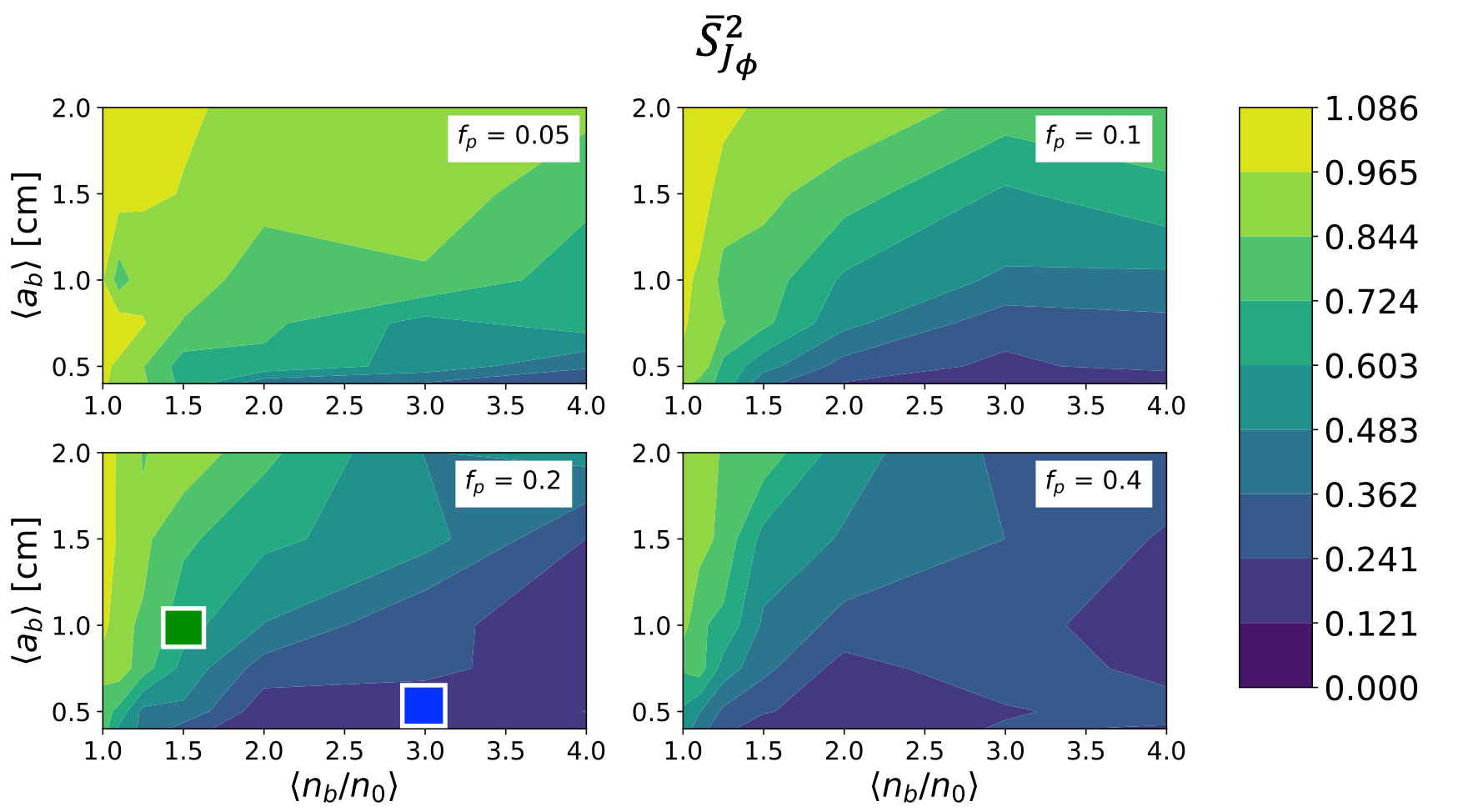}
\caption[font=5]{Parametric scan of turbulence parameters in GENRAY/CQL3D for discharge \# 1101104011. Plotted are contours of $\bar{S}^{2}_{J_{\phi}}$ (defined in Eq. (12) with $X$ replaced by the toroidal current drive profile). Green box: low turbulence case. Blue box: high turbulence case.}
\label{fig:5.15}
\end{figure}

A parametric scan of parameters $\langle n_b / n_0\rangle_{\text{grill}}$, $\langle a_b \rangle$, and $f_p$ is conducted. Figure \ref{fig:5.15} plots contours of $\bar{S}^{2}_{J_{\phi}}$. It is evident that $\bar{S}^{2}_{J_{\phi}}$ decreases as turbulence intensity increases. Here, we define ``turbulence intensity'' to mean the spatially averaged inverse scale length of density 
\begin{equation}
\langle \langle L^{-1}\rangle \rangle \equiv \langle \langle |\nabla n|/n_0 \rangle \rangle \approx f_p\langle n_b / n_0\rangle / \langle a_b \rangle
\end{equation}
where $\langle \langle ... \rangle \rangle$ denotes a spatial average. A minimum $\bar{S}^{2}_{J_{\phi}}\approx 0.15$ is achieved for $\langle \langle L^{-1}\rangle \rangle \approx 30 \,\text{m}^{-1}$, signifying a near seven-fold decrease in error. Interestingly, $\bar{S}^{2}_{J_{\phi}}$ remains near 0.15 for $\langle \langle L^{-1}\rangle \rangle \gtrsim 30 \,\text{m}^{-1}$. This suggests that the impact of scattering on LHCD saturates for sufficiently high turbulence intensity, and that this condition is met in experiment.

\begin{figure}[!h]
\centering
\includegraphics[width=16cm, height=8cm]{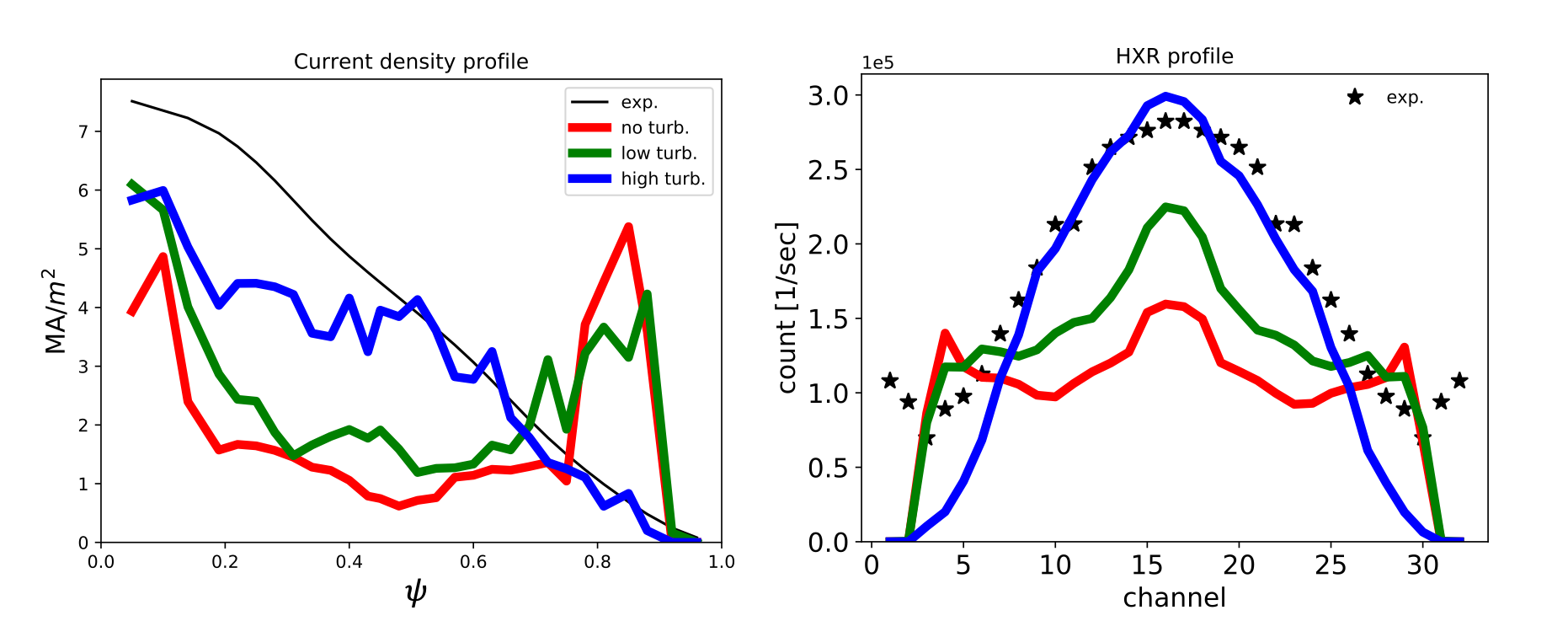}
\caption[font=5]{Current density (left) and HXR (right) profiles calculated in GENRAY/CQL3D for discharge \# 1101104011. Low and high turbulence correspond to points in Fig. \ref{fig:5.15}. Note: CQL3D top-down symmetrizes the plasma profile along the mid-plane, resulting in simulated HXR profiles roughly symmetric with respect to the middle channels (16 and 17). For proper comparison, the experimental HXR profile is also symmetrized.}
\label{fig:5.16}
\end{figure}

This is further illustrated by focusing on two points in the turbulence parameter space. The point $\left[ \langle n_b / n_0 \rangle_{\text{grill}}, \langle a_b \rangle, f_p \right] = [3, 0.5\,\text{cm},0.2]$ is termed ``high'' turbulence, as it results in saturated state (see blue box in Fig. \ref{fig:5.15}). The point $\left[ \langle n_b / n_0\rangle_{\text{grill}}, \langle a_b \rangle, f_p \right] = [1.5, 1\,\text{cm},0.2]$ is termed ``low'' turbulence, as $\bar{S}^{2}_{J_{\phi}} \approx 0.6$ and is therefore not fully saturated (see green box in Fig. \ref{fig:5.15}). Figure \ref{fig:5.16} plots the corresponding current density and HXR profiles. The reference case predicts a largely hollow current profile with peaks at $\psi \approx 0.1$ and 0.8. Correspondingly, the HXR profile is flat since this is a line-integrated diagnostic. These profiles are a poor match to the experiment. In contrast, the high turbulence case fully mitigates the off-axis peak in the current density profiles, and fills in the current ``valley'' in the $\psi \approx 0.2-0.7$ region. (Note that small fluctuations in the simulated ``high turb.'' current profile are a numerical artefact. They can be minimized by further increasing the number of rays simulated. In reality, fluctuations at this scale would be smoothed over by radial diffusion of fast electrons.) Consequently, the HXR profile is peaked in the middle channels and smoothly decays outwards. These profiles exhibit greatly improved agreement with experiment, although there remains a $\sim20\%$ underestimation of current density on-axis. The low turbulence case is qualitatively a mix of the saturated and reference case. Off-axis peaks are somewhat mitigated, and near-axis current is also somewhat increased compared to the reference case.

\subsection{Diffusion of ray-trajectories}
\begin{figure}[!h]
\centering
\includegraphics[width=12.5cm, height=8cm]{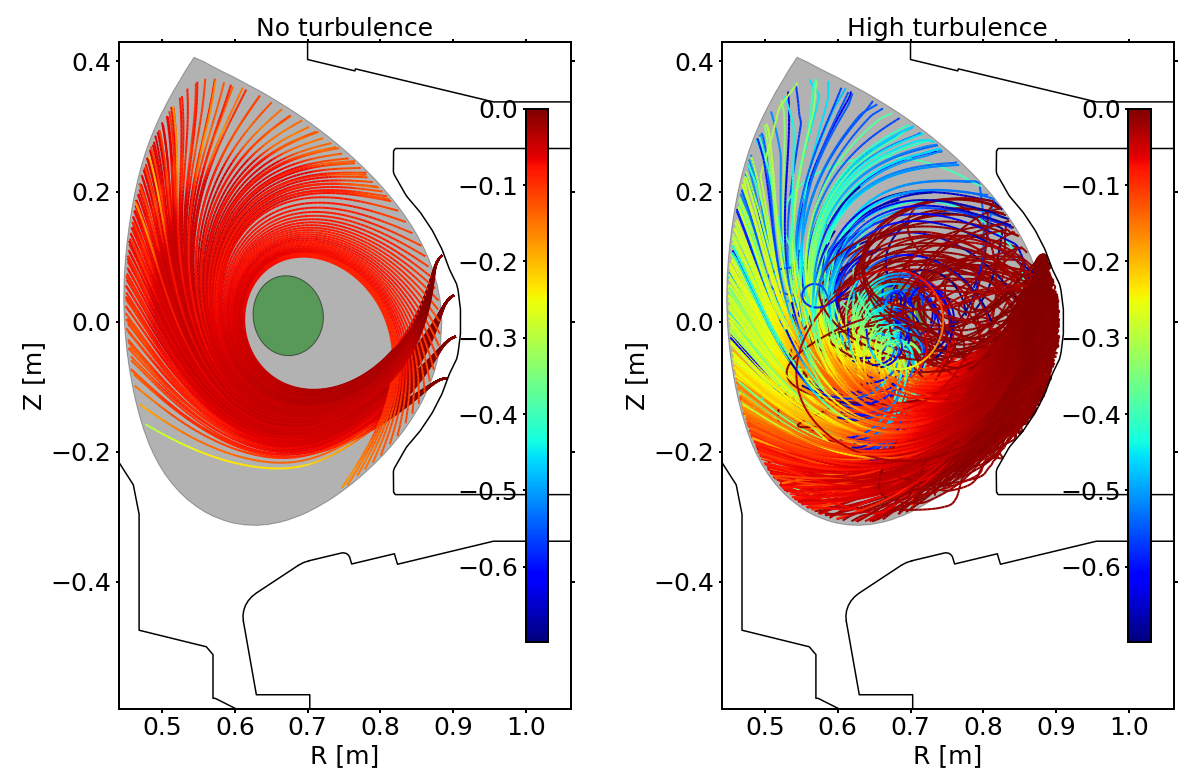}
\caption[font=5]{Ray-trajectories during first pass through the core. Plotting rays launched with $N_{||0} = 1.6 \pm 0.1$. Ray color denotes $\text{log}_{10}$ of normalized ray power. Green patch denotes region with $\psi \leq 0.2$.}
\label{fig:5.19}
\end{figure}

The scattering-induced diffusion of rays can be observed in both real- and phase-space. Figure \ref{fig:5.19} plots ray-trajectories in the poloidal plane during the first pass. In the reference case, ray damping in the core is negligible. In the high turbulence case, the initial ray bundle has greatly ``fanned'' due to scattering. It should be emphasized that in Fig. \ref{fig:5.19} is only plotting ray-trajectories during \emph{first-pass}, so what is shown is purely a scattering effect, and not due to ray-stochasticity following multiple reflections/passes. A fraction of rays scatter inwards, such that they strongly Landau damp in the hot near-axis region. This seeds a super-thermal electron tail near-axis, on which rays can further damp on subsequent passes. This explains the significant increase in near-axis current drive when scattering is included in the simulations. 

\begin{figure}[!h]
\centering
\includegraphics[width=15cm, height=7cm]{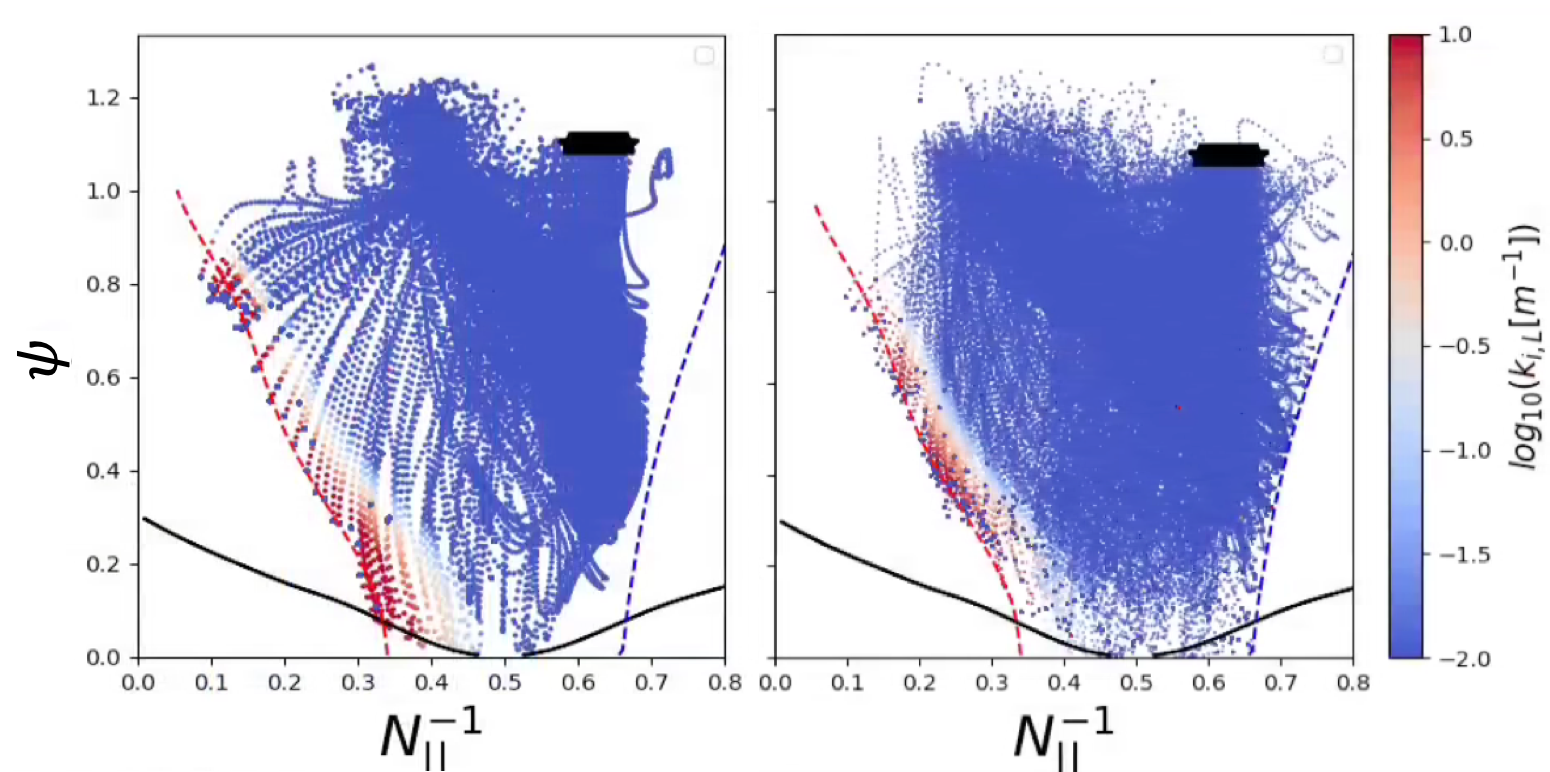}
\caption[font=5]{Ray-trajectories in phase-space for the reference (left) and high turbulence (right) cases. Plotting rays with $N_{||0}=1.6\pm0.1$. The thick black bar denotes start of rays, and ray color denotes imaginary component of wave-vector due to Landau damping. Red dashed like: strong linear ELD condition as defined in Eq. (1). Blue dashed like: the Stix-Golant accessibility condition. For completeness, we also plot the generalized accessibility condition \cite{takahashiGeneralizedAccessibilitySpectral1994,paolettiLowerHybridCurrent1994} in solid-black, though it is not a particularly relevant limit in C-Mod.}
\label{fig:5.22}
\end{figure}

Figure \ref{fig:5.22} plots ray-trajectories in phase-space for the reference and high turbulence case, showing only rays with $N_{||,0} = 1.6 \pm 0.1$. These rays are confined between the strong ELD contour (Eq. (1)) in dashed-red, and the Stix-Golant \cite{stixWavesPlasmas1992} accessibility contour in dashed-blue. After multiple passes, they eventually damp near the ELD contour. Two observations are made about the reference case. (1) Even though a wide sample of rays in $N_{||0}$ space is plotted, the allowable phase-space is only partially and non-uniformly filled in. (2) There is a distinct bundle of rays that strongly Landau damps at $\psi \approx 0.8$. This causes the off-axis peak in the current density profile in Figure \ref{fig:5.16}. Compare this to the high turbulence case, where phase-space is more uniformly filled in due to a strong diffusion of ray-trajectories caused by scattering. One particular result of this is the complete disappearance of the off-axis ray bundle. 

\subsection{High density partially inductive discharges}
Two higher density discharges, with partially non-inductive current drive, are modeled in GENRAY/CQL3D. 

In the first discharge (\#1140411016), 460kW of LH power launched at $N_{||0}=1.9$ is coupled to an L-mode plasma with $\bar{n}_{e}= 0.8\times 10^{20} \,\text{m}^{-3}$ and $T_{e0} = 2.2\,$keV. During the LH phase, $V_{\text{loop}}$ decreases from 1V to 0.4V.  In the second discharge (\#1140411015), 450kW of LH power launched at $N_{||0}=1.9$ is coupled to plasma with $\bar{n}_{e}=1.1\times 10^{20}\,\text{m}^{-3}$ and $T_{e0}=1.8\,$keV. Correspondingly, loop voltage drops from 0.95V to 0.75V. Note that this plasma is above the LHCD density limit \cite{wallaceLowerHybridCurrent2011}, which explains the relatively small change in loop voltage. The DC electric field is neglected during CQL3D modeling of both discharges. There remain large uncertainties about the DC field radial profile, which can significantly modify the current profile. This can be accurately accounted for using time-dependent simulations as done by Poli \emph{et al.}, (2016) \cite{poliExperimentalModelingUncertainties2016}.

\comment{
The DC electric field is neglected during CQL3D modeling. There are two reasons for this decision: First, there remain uncertainties about the spatial profile of the DC electric field, which can modify the total current profile. Second, there exist the issue of a ``slide-away'' electron population excited in CQL3D if there is both RF power and a sufficiently large DC field.}

\begin{figure}[!h]
\centering
\includegraphics[width=16cm, height=8cm]{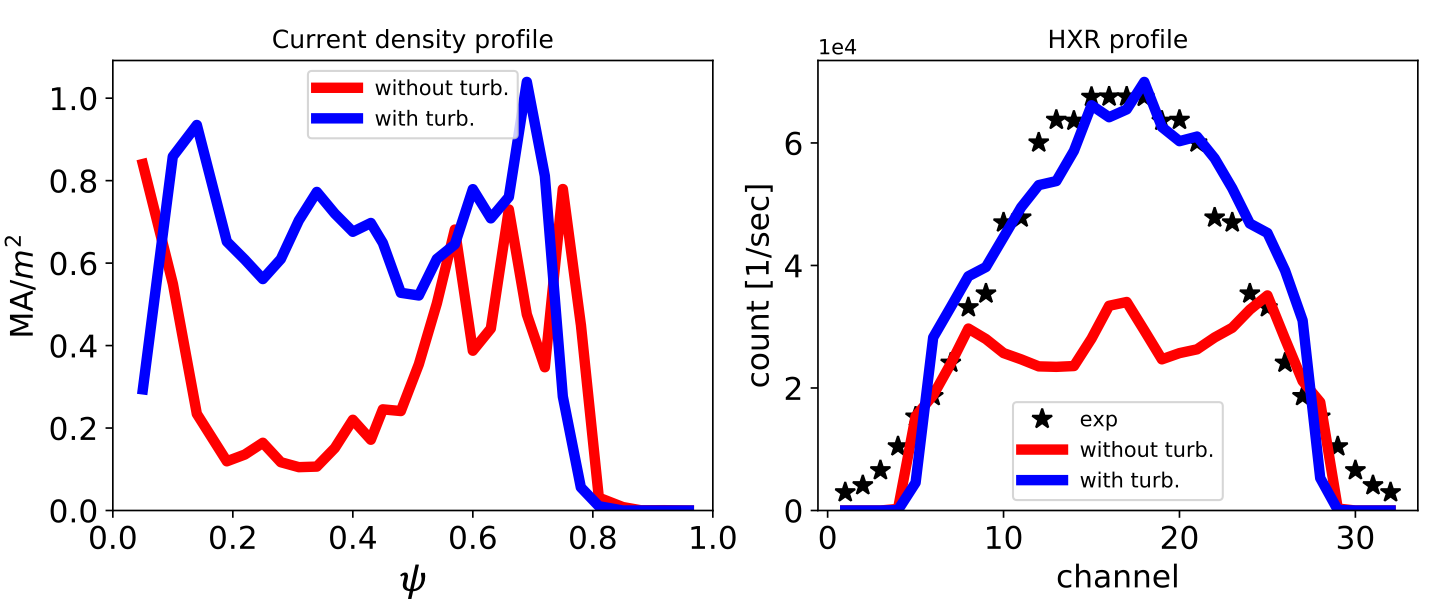}
\caption[font=5]{Current density (left) and HXR (right) profiles calculated in GENRAY/CQL3D for discharge \#1140411016 for which $\bar{n}_{e}= 0.8\times 10^{20} \,\text{m}^{-3}$, $T_{e0} = 2.2\,$keV, and $\left[ \langle n_b / n_0 \rangle_{\text{grill}}, \langle a_b \rangle, f_p \right] = [3, 0.5\,\text{cm},0.2]$.}
\label{fig:5.27}
\end{figure}

\begin{figure}[!h]
\centering
\includegraphics[width=16cm, height=8cm]{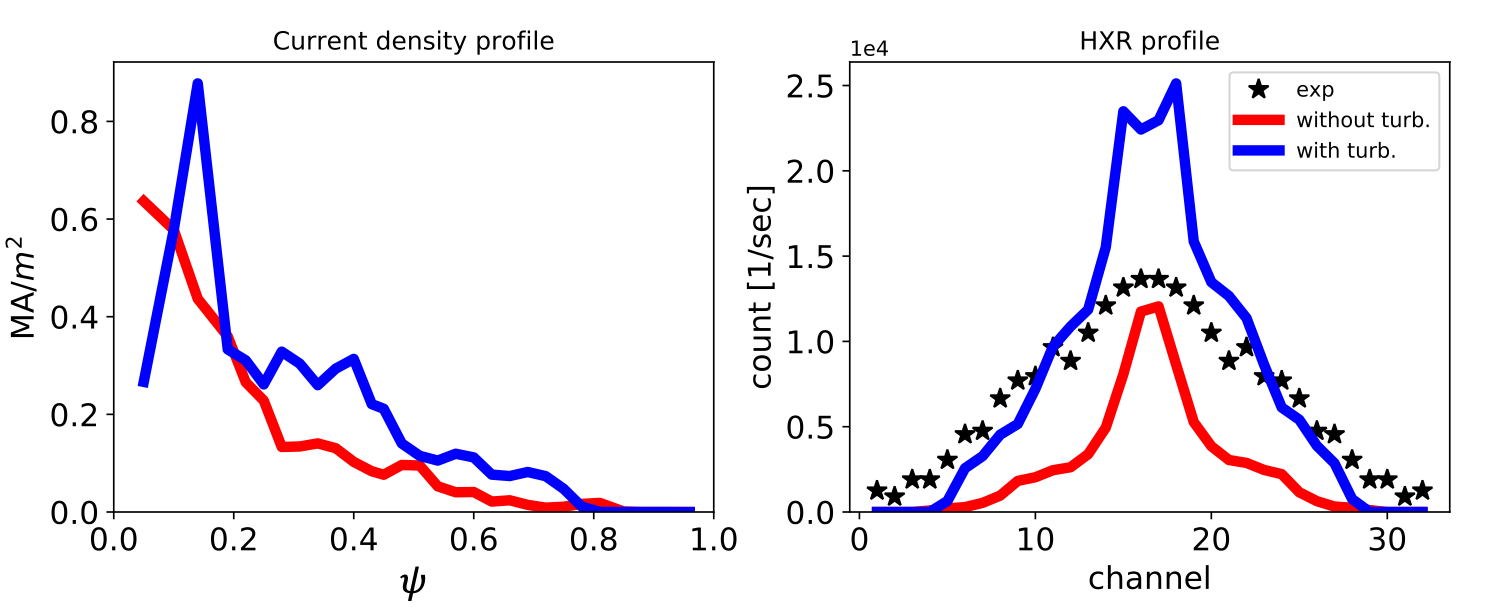}
\caption[font=5]{Current density (left) and HXR (right) profiles calculated in GENRAY/CQL3D for discharge \#1140411015 for which $\bar{n}_{e}=1.1\times 10^{20}\,\text{m}^{-3}$, $T_{e0}=1.8\,$keV, and $\left[ \langle n_b / n_0 \rangle_{\text{grill}}, \langle a_b \rangle, f_p \right] = [3, 0.5\,\text{cm},0.2]$.}
\label{fig:5.32}
\end{figure}

Figures \ref{fig:5.27} and \ref{fig:5.32} plot the simulated current density and HXR profiles for the $\bar{n}_{e}= 0.8$ and $1.1 \times 10^{20} \,\text{m}^{-3}$ discharges, respectively. The cases with turbulence use the same filament parameters as the high turbulence case identified in Section 3.2. No attempt is made to compare to experimental current densities. MSE measurements were not available for these discharges to reconstruct the current profile. Even if it were available, it is difficult to calculate the Ohmic and non-inductive contributions separately because the DC electric field profile is not known.

Notably, for the $\bar{n}_{e}= 0.8 \times 10^{20} \,\text{m}^{-3}$ discharge, the current valley at $\psi \approx 0.3$ disappears when scattering is modelled. For both discharges, turbulence results in greater current drive near-axis ($\psi < 0.4$), ultimately leading to improved agreement with HXR profile measurements. 

\begin{figure}[!h]
\centering
\includegraphics[width=16cm, height=10cm]{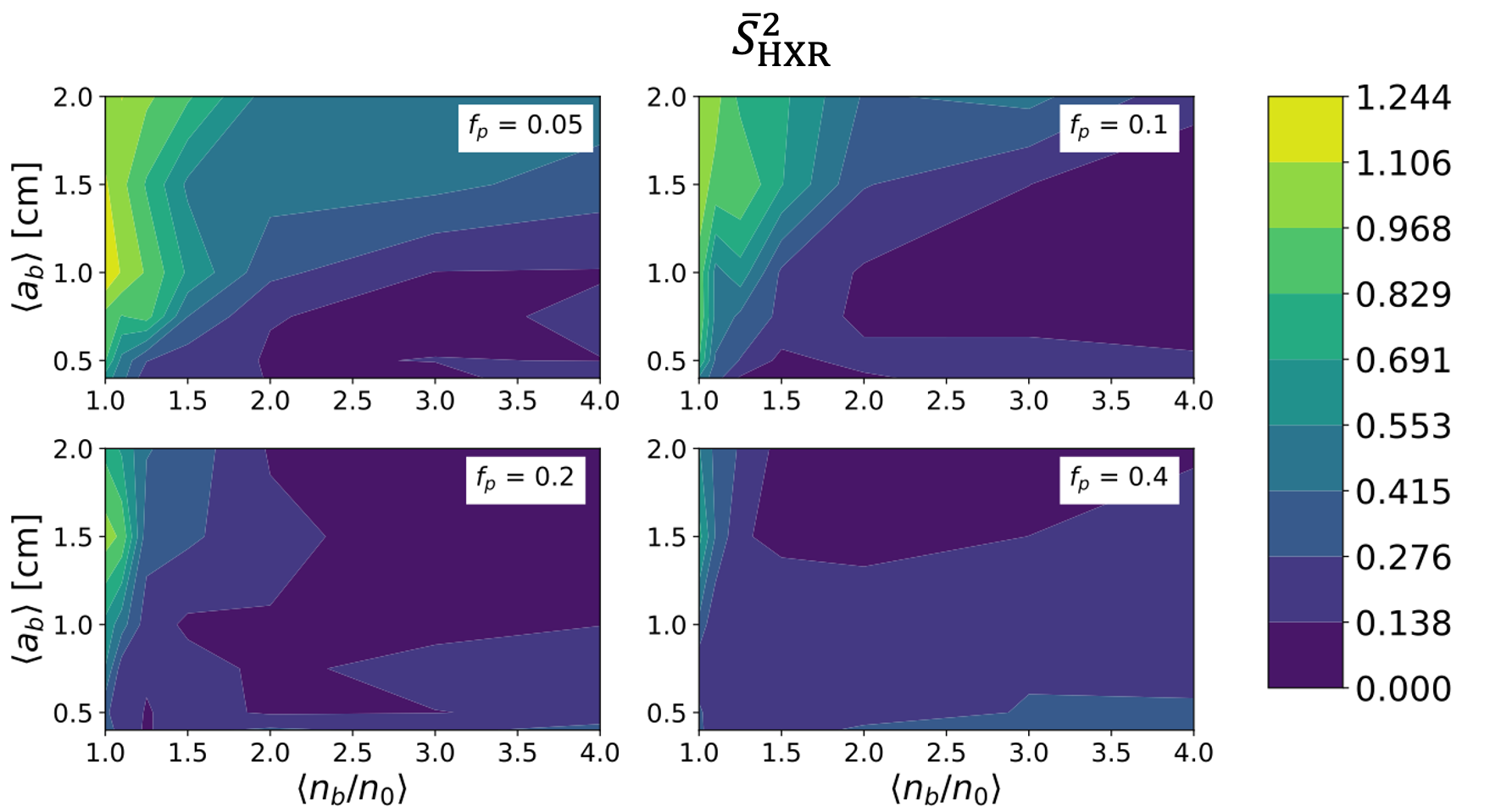}
\caption[font=5]{Parametric scan of turbulence parameters in GENRAY/CQL3D for discharge \#1140411015. Plotted are contours of $\bar{S}^{2}_{\text{HXR}}$ (defined in Eq. (12) with $X$ replaced by the HXR profile).}
\label{fig:5.34}
\end{figure}

Similar to the low-density fully non-inductive case, a scan of filaments parameters is conducted for the $\bar{n}_{e}= 0.8 \times 10^{20} \,\text{m}^{-3}$ discharge in Fig. \ref{fig:5.34}. Given that the experimental current drive profile cannot be isolated, we instead plot $\bar{S}^{2}_{\text{HXR}}$, for which the array $X$ in Eq. (12) is replaced with count rates from the 32-channel HXR diagnostic. It is evident that the saturation effect of SOL turbulence on LHCD is also present at high densities, in this case for $\langle \langle L^{-1}\rangle \rangle \gtrsim 15 \text{m}^{-1}$. Thus, it is confirmed that the high turbulence parameters identified as being in the saturated regime for the low density ($\bar{n}_{e} = 0.52 \times 10^{20} \,\text{m}^{-3}$) case, is also in the saturated regime for higher density discharges.

\subsection{Impact of scattering asymmetry}
As mentioned in Sec. 3.1, the dominant asymmetric scattering effect is $S \rightarrow S$ scatter. On average, this leads to rays rotating away from the core. In turn, these rays propagate longer distances in the SOL, where parasitic collisional damping is non-negligible. One might then expect that less current is driven when asymmetric scatter is accounted for. A simple gedanken experiment is conducted to test this hypothesis. Three cases are run: (1) the reference case, (2) a case with the usual scattering model, and (3) a case with the scattering model assuming reversed parity. Here, a reversal of parity means setting $\sigma_{\text{eff},j \rightarrow j'}(\chi) \rightarrow \sigma_{\text{eff},j \rightarrow j'}(-\chi)$, such that, on average, rays scatter \emph{towards} the core. 

\begin{figure}[!h]
\centering
\includegraphics[width=15cm, height=7.5cm]{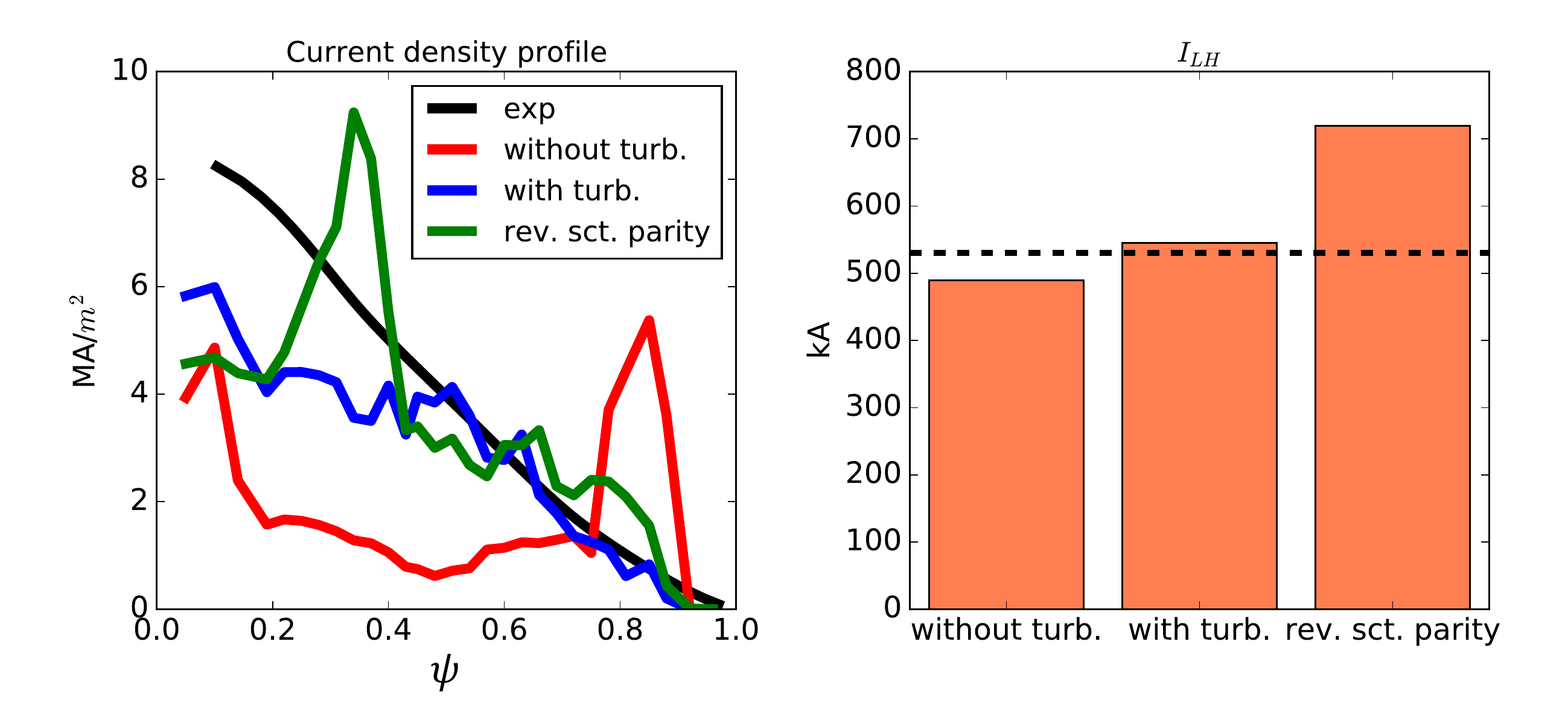}
\caption[font=5]{Current density profiles calculated in GENRAY/CQL3D for discharge \#1101104011. Testing a reversed scattering parity. Horizontal dashed line denotes experimental value.}
\label{fig:5.36}
\end{figure}

\begin{figure}[!h]
\centering
\includegraphics[width=15cm, height=7.5cm]{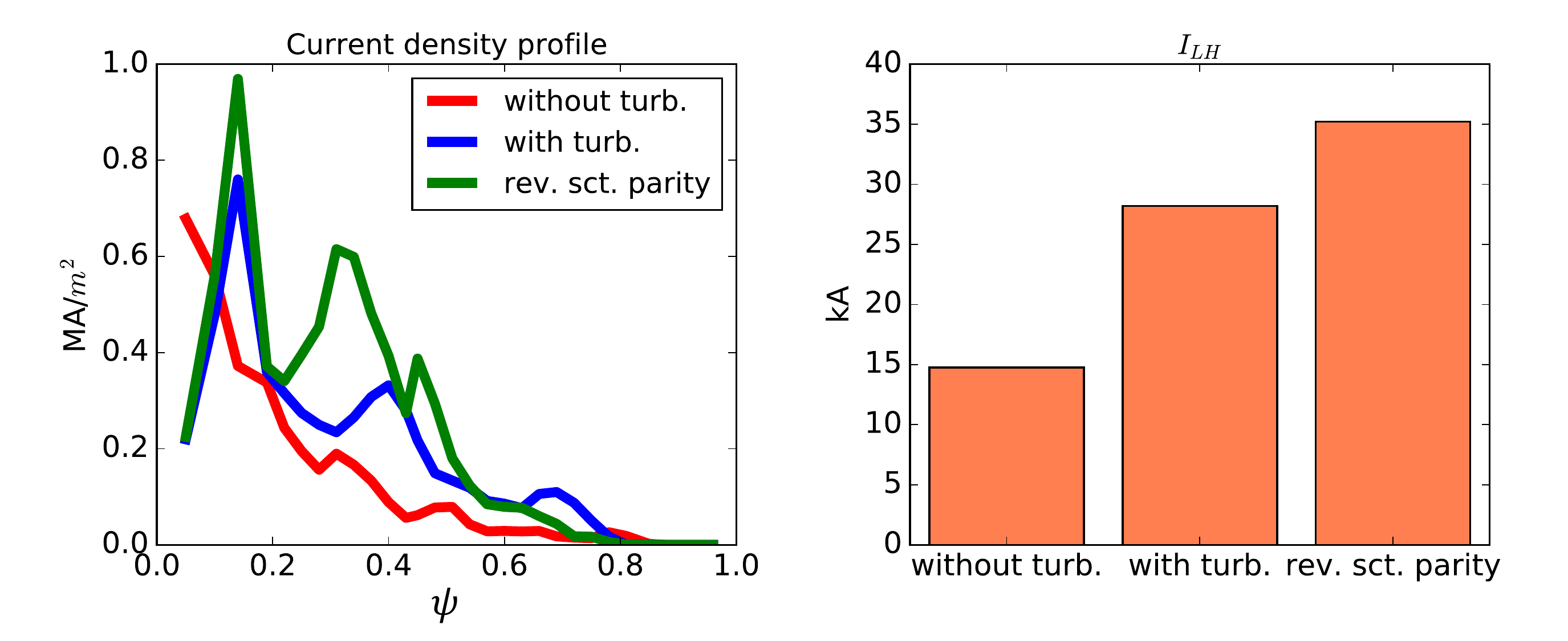}
\caption[font=5]{Current density profiles calculated in GENRAY/CQL3D for discharge \#1140411015. Testing a reversed scattering parity. Simulated current profile not compared to experimental measurements because the Ohmic contribution is neglected in the model.}
\label{fig:5.38}
\end{figure}

First, this gedanken experiment is conducted for the low density fully non-inductive C-Mod discharge. Figure \ref{fig:5.36} plots resulting current density profiles. It is revealed that flipping the scattering parity does indeed affect the profile. In the reverse parity case, a large current density peak forms at $\psi \approx 0.2-04$. Relative to the normal scattering case, the reverse parity case results in $\sim30\%$ greater current drive. This trend is as expected. This experiment is repeated with the $\bar{n}_{e} = 1.1 \times 10^{20} \,\text{m}^{-3}$ discharge, with Figure \ref{fig:5.38} plotting current density profiles. A reversed scattering parity results in significantly greater near-axis current deposition, and $\sim25\%$ greater total current drive compared to the normal scattering case.

Contrary to the hypothesis, the normal parity case results in greater current than in the reference case. An explanation for this can be found in examining Fig. \ref{fig:5.19}. While there is a net rotation of rays away from the core, there is still a large fraction of rays rotated towards the core such that they damp on-axis during first pass. In contrast, virtually no power damps on-axis during first pass in the reference case. Thus, this effect is enough to counteract the net outward scatter.

\section{Summary and discussion}
A multiscale model has been developed for LH wave propagation in a magnetized plasma with filamentary turbulence. Wave propagation is computed using ray-tracing. Relatively expensive full-wave calculations are reserved only for wave-filament interactions, which then act as stochastic kicks to the ray-trajectories. The scattering probabilities are spatially dependent, which allows the prescription of a radially and poloidally tapered SOL turbulence profile in a tokamak geometry. In addition, both like- and unlike-mode scattering is modelled.

This multiscale scattering model is applied to C-Mod discharges, where the impact of scattering on LHCD is shown to saturate at sufficiently high levels of turbulence intensity. In the saturated phase, ray-trajectories are significantly diffused in phase-space.  In turn, LHCD predictions are insensitive to the assumed turbulence parameters. At low density ($\bar{n}_{e} \approx 0.52 \times 10^{20}\,\text{m}^{-3}$), the saturated case results in excellent agreement with experimental MSE and HXR measurements. Here, a turbulence intensity of $\langle \langle L^{-1}\rangle \rangle \gtrsim 30 \,\text{m}^{-1}$ resulted in saturation with $\bar{S}^2_{J_{\phi}}<0.15$, signifying a near seven-fold decrease in error for the current profile (compared to the simulation with no scattering). This saturation is also present in the high density case above the LHCD density limit. Here, saturation at $\langle \langle L^{-1}\rangle \rangle \gtrsim 15 \,\text{m}^{-1}$ results in significantly improved match to experiment ($\bar{S}^2_{\text{HXR}}<0.1$). This suggests that, experimentally, the impact of wave-filament scatter is saturated in C-Mod L-mode discharges for all background densities. In future work, it will be worthwhile to study how the threshold $\langle \langle L^{-1}\rangle \rangle$ depends on plasma and LH launch parameters.

While this modeling suggests that C-Mod LH launch is in this saturation regime, it is unclear whether this is corroborated by edge turbulence measurements. Analysis of Langmuir probe measurements in the SOL of C-Mod measure $\langle n_b/n_0\rangle \approx 2.1$ \cite{kubeFluctuationStatisticsScrapeoff2016}. Similarly, GPI measurements indicate $\langle a_b \rangle \approx1\,$cm \cite{kubeBlobSizesVelocities2013} and $f_{p} \sim 0.05$ \cite{zwebenEstimateConvectiveRadial2011}. However, these burst statistic analyses require setting an arbitrary threshold above r.m.s. values for identifying a filament. This threshold affects the computed $\langle n_b/n_0\rangle$, $\langle a_b \rangle$, and greatly affects $f_{p}$. Given the large experimental uncertainties in measuring these parameters, no attempt is made here in prescribing the likeliest filament PDF or packing fraction. Rather, it is argued that SOL measurements in C-Mod do not rule out the possibility of a saturated regime for LH wave scattering.

The impact of asymmetric scattering was tested by reversing the scattering parity. The reversed parity leads to increased scatter of LH power into the core. As expected, this results in greater current drive and HXR count compared to simulations with the correct scattering parity. It is important to note, however, that scatter with correct parity, which results in net outward scatter, still leads to greater current drive than the case with no scattering. This is attributed to the non-zero fraction of LH power that scatters inwards and Landau damps near-axis on first pass.

In regards to the three C-Mod discharges modelled in this work, it is interesting to note that the simulated HXR count, for both the reference and high turbulence cases, agree within an order of magnitude with the experimental values. This is in stark contrast to previous analysis, in which standard ray-tracing/Fokker-Planck simulations could not replicate the anomalous drop in total current driven and HXR count for $\bar{n}_{e}\gtrsim 10^{20}\,\text{m}^{-3}$ \cite{wallaceLowerHybridCurrent2011,mumgaardLowerHybridCurrent2015}. However, later simulations including a more realistic collisional SOL could successfully do so \cite{shiraiwaImpactSOLPlasma2015}. This present work uses this improved SOL model. Overall, this suggests that proper modeling of SOL collisional losses is needed to accurately calculate \emph{total} current and HXR count, while proper modeling of scattering is needed to accurately calculate the \emph{radial profile} of current and HXR count.

Lastly, it should be emphasized that these multiscale simulations allow, for the first time, the coupling of wave-filament scattering physics, beyond ray-tracing, to the down-stream physics of core damping and current drive. This is of practical importance for the following reason. It has been shown, through the use of this multiscale model, that accurate simulation of LHCD requires accounting for wave-filament scatter from sufficiently intense SOL turbulence. This corresponds to a situation in which ray-tracing is invalid, and so higher-order scattering effects must be taken into account. This is evidenced by the relatively poor performance of the filament refraction model which employed ray-tracing alone \cite{biswasStudyTurbulenceinducedRefraction2020}. While global full-wave simulations are computationally in-feasible, this multiscale approach applies full-wave calculations only locally where it is needed (at the wave-filament interaction). In this way, fast parametric scans and intershot analysis is made possible on modest, university-scale computing clusters. For example, consider Fig. \ref{fig:5.15}. This parametric scan consisted of 196 simulations and required $\sim 3500$ total CPU-hours on the Engaging cluster. (Equivalently, $\sim17.5$ CPU-hours were required per simulation.)

\section*{Acknowledgements}
This work was supported by US DoE under contract numbers: DE-SC0018090 supporting the RF-SciDAC 4 project, DE-SC0014264 supporting PSFC MFE projects, and DE-FC02-04ER54698 supporting DIII-D projects. The GENRAY/CQL3D and PETRA-M simulations presented in this paper were performed on the MIT-PSFC partition of the Engaging cluster at the MGHPCC facility (www.mghpcc.org) which was funded by DoE grant number DE-FG02-91-ER54109.

\section*{Declaration of interest}
The authors report no conflict of interest.

\appendix
\setcounter{equation}{0}
\renewcommand{\theequation}{A.\arabic{equation}}\renewcommand{\thesubsection}{\Alph{subsection}}
\section{Radially and poloidally tapered turbulence}
GENRAY uses $(\psi, \theta_{p})$-coordinates in the poloidal plane, where $\psi \equiv \sqrt{\Psi_{\phi}/\Psi_{\phi, a}}$ is the normalized radial coordinate (and $\Psi_{\phi}$ is toroidal magnetic flux). The evaluation of coordinate $\psi$ is approximately extended outside the separatrix. It is approximated that $n(\bold{r}) \approx n(\psi,\theta_{p})$. Given the relatively slow spatial variation of $B$, it is acceptable for our purposes of calculating scattering probabilities in the SOL to assume
\begin{equation}
B \approx B_{\phi} \approx \frac{B_{0}}{1 + \frac{a}{R_0} \cos(\theta_{p})}
\end{equation}
where $B_{\phi}$ is the toroidal magnetic field and $a,R_0$ are the minor, major plasma radius.

In assuming that the turbulence parameters vary radially and poloidally (i.e. $p(n_b/n_0(\bold{r}),a_b(\bold{r})) = p(n_b/n_0, a_b; \psi, \theta_{p})$), one can parametrize the scattering probabilities as $\hat{\sigma}_{\text{eff},j \rightarrow j'}(\chi; \psi, \theta_{p}, k_{||})$ and $\Sigma_{\text{eff},j \rightarrow j'}(\chi; \psi, \theta_{p}, k_{||})$. These form the lookup tables in GENRAY.

As in \cite{biswasHybridFullwaveMarkov2021}, it is assumed that the filament joint-PDF is a bivariate skewed Gaussion\cite{azzaliniStatisticalApplicationsMultivariate1999}, now with radial and poloidal dependence. Its analytic form is written as a function of the 2D column vector $\zeta = \left[ n_b/n_0; a_b \right]$, while parametrization depending on radial and poloidal coordinates is made explicit. 
\begin{subequations}
\begin{align}
p(\zeta;\psi,\theta_{p}) &= 2 \phi_{2}(\zeta - Q(\psi,\theta_{p})\Omega(\psi,\theta_{p})) \Phi(\Gamma^{\text{T}}(\psi,\theta_{p}) \left[ \zeta-Q(\psi,\theta_{p})\right])\\
\Omega(\psi,\theta_{p}) & =
\begin{bmatrix}
 s_{n_b/n_0}^{2}(\psi,\theta_{p}) & \eta s_{n_b/n_0}(\psi,\theta_{p}) s_{a_b}\\
 \eta s_{n_b/n_0}(\psi,\theta_{p}) s_{a_b} & s_{a_b}^{2}
\end{bmatrix}\\
Q(\psi,\theta_{p}) & = \langle \zeta \rangle(\psi,\theta_{p}) - \sqrt{\frac{2}{\pi}} \delta(\psi,\theta_{p})\\
\delta(\psi,\theta_{p}) & = \frac{1}{\sqrt{1 + \Gamma^{\text{T}} \Omega(\psi,\theta_{p}) \Gamma}} \Omega(\psi,\theta_{p}) \Gamma\\
\Gamma(\psi,\theta_{p}) & =
\begin{bmatrix}
\Gamma_{n_b/n_0}(\psi,\theta_{p})\\
\Gamma_{a_b}
\end{bmatrix}
\end{align}
\end{subequations}
where $\phi_{2}(\zeta,\Omega)$ is a 2D Gaussian PDF with zero mean and correlation matrix $\Omega$, $\Phi(x)$ is the 1D Gaussian CDF for scalar input $x$, $\Gamma$ is now a 2D column vector of skewness factors, and $\langle \zeta \rangle\equiv \left[ \langle n_b/n_0 \rangle; \langle a_b\rangle \right]$ is the 2D column vector of mean $n_b/n_0$ and $a_b$. Similarly, $s \equiv [s_{n_b/n_0}; s_{a_b}]$ are standard deviations, and $\eta$ is the scalar correlation coefficient. It is therefore possible to parameterize $p(\zeta)$ as a function of  $\langle \zeta \rangle, \Gamma, s$, $\eta$, $\psi$, and $\theta_{p}$. In contrast to \cite{biswasHybridFullwaveMarkov2021}, the filament relative density mean ($\langle n_{b}/n_{0}\rangle$), standard deviation ($s_{n_{b}/n_{0}}$), and skewness ($\Gamma_{n_b/n_0}$) are allowed to vary with $\psi$ and $\theta_{p}$. For simplicity, the filament width statistics ($\langle a_{b} \rangle$, $s_{a_{b}}$, and $\Gamma_{a_{b}}$) are kept homogeneous. Further, it is assumed that $s_{a_b} = 0.2\,$cm and $\Gamma_{a_{b}} = 7$. The correlation coefficient $\eta = 0.9$, signifying strong positive correlation between $n_{b}/n_{0}$ and $a_b$ \cite{decristoforoBlobInteractions2D2020}.

The radial profile for fluctuations should smoothly increase from zero inside the seperatrix to the user-specified $\langle \frac{n_{b}}{n_{0}}\rangle_{\text{grill}}$ at the LH grill. The following analytic forms are assumed for $\langle n_{b}/n_{0}\rangle$, $s_{n_{b}/n_{0}}$ and $\Gamma_{n_b/n_0}$:
\begin{subequations}
\begin{align}
\langle n_{b}/n_{0}\rangle & = \left[ \langle n_{b}/n_{0}\rangle_{\text{min}} +  (\langle n_{b}/n_{0}\rangle_{\text{grill}} - \langle n_{b}/n_{0}\rangle_{\text{min}}) \bar{\psi} - 1\right]^{g(\theta_{p})} +1\\
s_{n_{b}/n_{0}} & = \left[ s_{n_{b}/n_{0},\text{min}} +  (s_{n_{b}/n_{0},\text{grill}} - s_{n_{b}/n_{0},\text{min}}) \bar{\psi}\right]^{g(\theta_{p})}\\
\Gamma_{n_{b}/n_{0}} & = \left[ \Gamma_{n_{b}/n_{0},\text{min}} +  (\Gamma_{n_{b}/n_{0},\text{grill}} - \Gamma_{n_{b}/n_{0},\text{min}}) \bar{\psi}\right]^{g(\theta_{p})}
\end{align}
\end{subequations}
where 
 \begin{equation}
\bar{\psi}(\psi) = \begin{cases}
0 & \psi < \psi_{\text{min}}\\
(\psi-\psi_{\text{min}})/(\psi_{\text{grill}}-\psi_{\text{min}}) &  \psi_{\text{min}}<\psi < \psi_{\text{grill}}\\
1 &  \psi > \psi_{\text{grill}}
\end{cases}
\end{equation}
and $\psi_{\text{grill}}$ is the radial location of the LH launcher (a.k.a. grill). In this study, the minimum radial location of non-zero turbulence ($\psi_{\text{min}}$) is set to 0.85. It is assumed that $\langle n_{b}/n_{0}\rangle_{\text{min}} = 1$, $s_{n_{b}/n_{0},\text{min}} = 0.1$, $\Gamma_{n_{b}/n_{0},\text{min}} = 0.01$, and $\Gamma_{n_{b}/n_{0},\text{max}} = 7$. The following linear relation is used to prescribe $s_{n_{b}/n_{0},\text{grill}}$:
\begin{equation}
s_{n_{b}/n_{0},\text{grill}} = c_{1} \left\langle \frac{n_{b}}{n_{0}} \right\rangle_{\text{grill}} + c_{2}
\end{equation}
where $c_{1}$ and $c_{2}$ are set such that $s_{n_{b}/n_{0},\text{grill}}=1.1$ for $\langle \frac{n_{b}}{n_{0}}\rangle_{\text{grill}}=4$ and $s_{n_{b}/n_{0},\text{grill}}=0.22$ for $\langle \frac{n_{b}}{n_{0}}\rangle_{\text{grill}}=1.1$.

Lastly, the function $g(\theta_{p})$ introduces a poloidal variation and is prescribed to be
\begin{equation}
g(\theta_{p}) = 0.1 + 0.9 \text{cos}^{2}(\theta_{p} /2)
\end{equation}
This accounts for experimental measurements that density fluctuations on the HFS are greatly suppressed \cite{terryObservationsTurbulenceScrapeofflayer2003,smickPlasmaProfilesFlows2005}. 

\begin{figure}[!h]
\centering
\includegraphics[width=9cm, height=8cm]{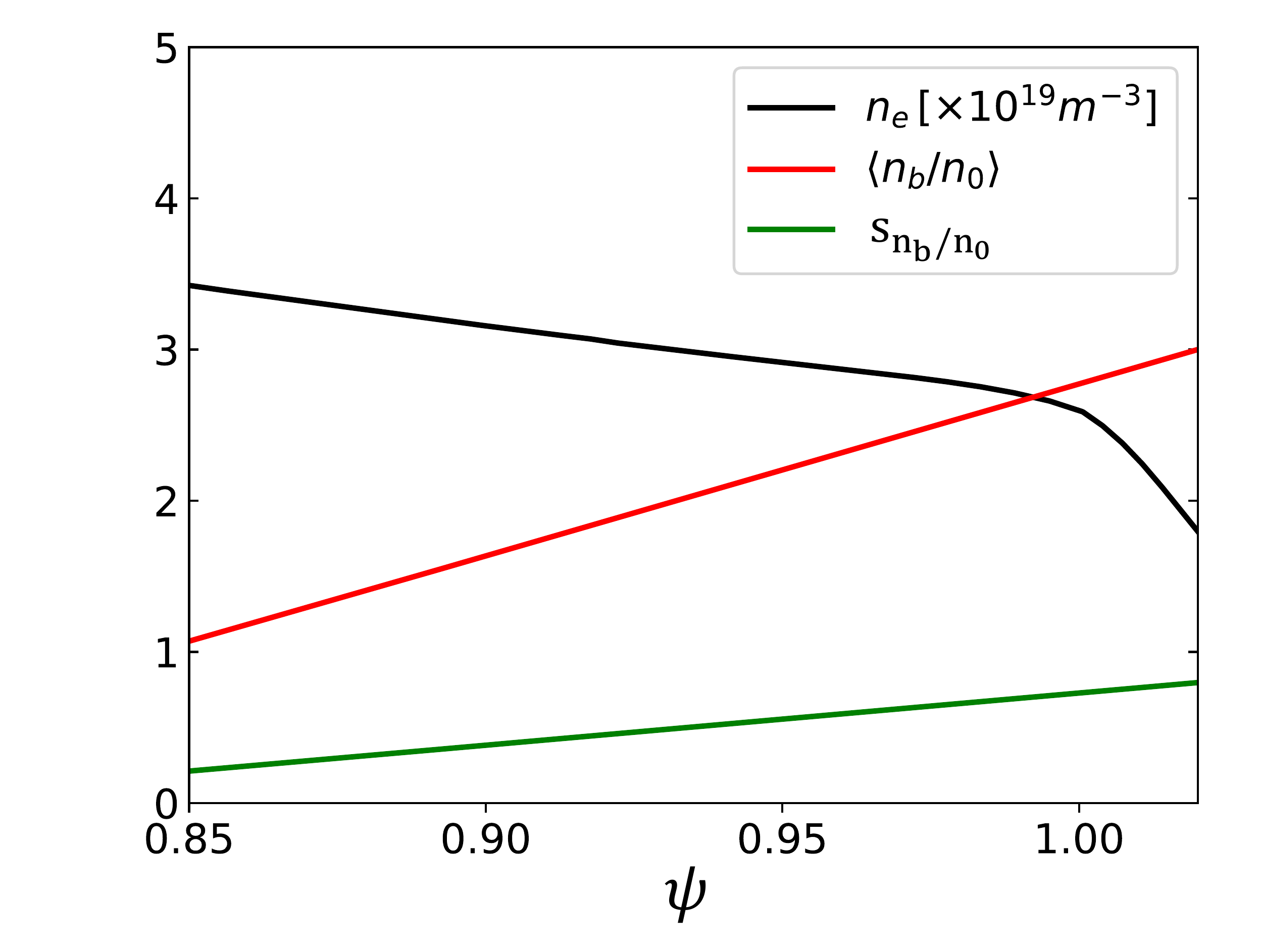}
\caption[font=5]{Background density and modeled turbulence profile at the outer mid-plane in a C-Mod discharge. The only user-specified parameter is $\langle n_{0}/n_{b}\rangle_{\text{grill}} = 3$. The $\langle n_{0}/n_{b}\rangle$ and $s_{n_{b}/n_{0}}$ profiles are prescribed according to Eq. (A.3-A.5). }
\label{fig:app1}
\end{figure}

Figure \ref{fig:app1} plots background density and example turbulence profiles at the outer mid-plane for the non-inductive C-Mod discharge. In this case, the LH grill position corresponds to $\psi_{\text{grill}} \approx 1.02$. The free-parameter $\langle n_{0}/n_{b}\rangle_{\text{grill}} = 3$ prescribes mean relative density of filaments at the LH grill. The mean relative density of filaments monotonically decrease to unity at $\psi=0.85$, meaning they are indistinguishable from the background, thereby smoothly transitioning into a quiescent core plasma.

\begin{figure}[!h]
\centering
\includegraphics[width=16cm, height=8cm]{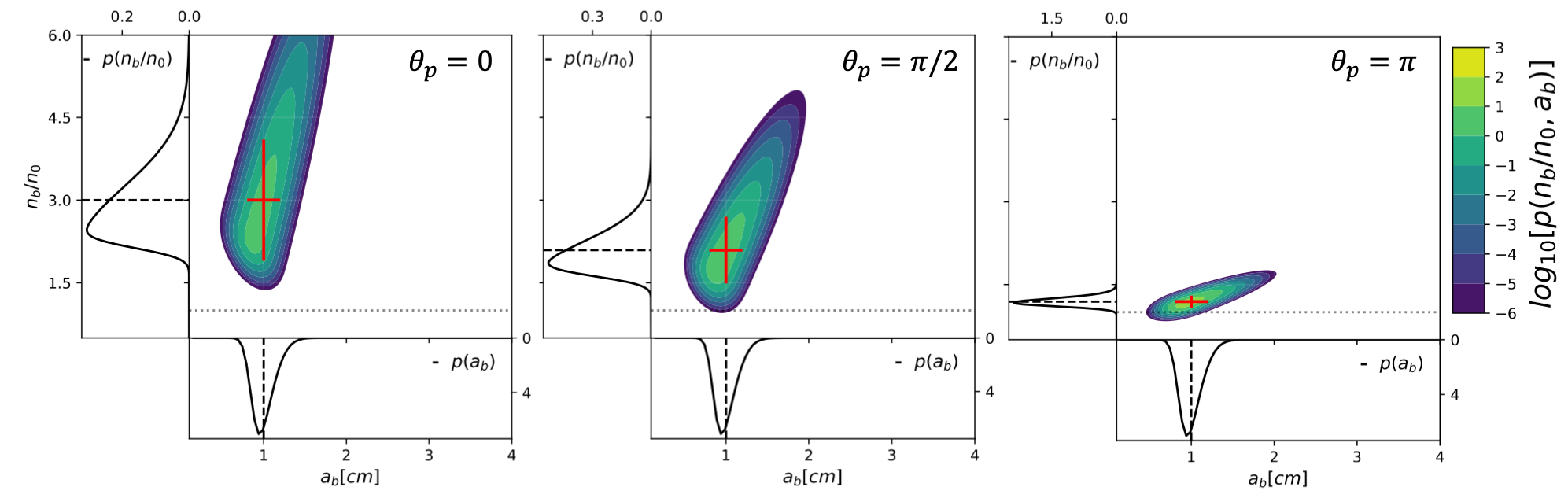}
\caption[font=5]{Joint-PDF of filament parameters for $\psi=1.02$. The only user-specified parameters are $\langle n_{0}/n_{b}\rangle_{\text{grill}} = 3$ and $\langle a_b \rangle = 1\,$cm. Red cross-hairs denote $\pm$ 1 standard deviation in $n_{b}/n_{0}$ and $a_{b}$.}
\label{fig:app2}
\end{figure}

Figure \ref{fig:app2} shows an example of how the joint-PDF at $\psi = 1.02$ (in the SOL) is tapered poloidally such that filaments are, on average, less dense on the high-field side.

\clearpage
\bibliographystyle{ieeetr}
\bibliography{./paper_refs}
\end{document}